\documentclass[%
 aip,
 amsmath,amssymb,
 floatfix,  
 reprint,%
]{revtex4-1}

\usepackage{graphicx}
\usepackage{dcolumn}
\usepackage{bm}
\usepackage[utf8]{inputenc}
\usepackage[T1]{fontenc}
\usepackage{mathptmx}
\usepackage{etoolbox}
\usepackage{xcolor}
\makeatletter
\def\@email#1#2{%
 \endgroup
 \patchcmd{\titleblock@produce}
  {\frontmatter@RRAPformat}
  {\frontmatter@RRAPformat{\produce@RRAP{*#1\href{mailto:#2}{#2}}}\frontmatter@RRAPformat}
  {}{}
}%
\makeatother
\begin{document}

\preprint{AIP/123-QED}

\title[Observation of Long-Lifetime Magnon Pairs by Fano Resonance of Photons]{Observation of Long-Lifetime Magnon Pairs \\by Fano Resonance of Photons}
\author{Qian-Nan Huang}
\affiliation{School of Physics, Huazhong University of Science and Technology, Wuhan 430074, China
}%
\author{Zhiping Xue}%
 \affiliation{School of Physics, Huazhong University of Science and Technology, Wuhan 430074, China
}%

\author{Tao Yu}
\email{taoyuphy@hust.edu.cn}
\affiliation{School of Physics, Huazhong University of Science and Technology, Wuhan 430074, China
}%

\date{\today}

\begin{abstract}
Mode fluctuations with a long lifetime are essential for quantum information and logic operations in magnonic devices.
We probe the broadband nonlinear magnetization dynamics of a high-quality ferromagnet under a strong microwave drive using microwave spectroscopy. We observe an \textit{unexpected} Fano resonance in the microwave transmission when the driven amplitude of the magnetization is large and the drive frequency $\omega_d$ is close to but not at the ferromagnetic resonance. We interpret this Fano resonance by a scattering theory of photons considering the three-magnon interaction between the Kittel magnon and magnon pairs with opposite wave vectors of frequency $\omega_d/2$. The theoretical model suggests that the microwave spectroscopy measures the dynamics of the fluctuation $\delta \hat{\alpha}$ of the Kittel magnon and $\delta\hat{\beta}_{\pm k}$ of the magnon pairs over the driven steady states, which are coupled coherently by the steady-state amplitudes. With the damping of $\delta\hat{\beta}_{\pm k}$ much smaller than that of $\delta \hat{\alpha}$, the theoretical calculation well reproduces the observed Fano resonance, indicating the magnon pairs hold a recorded long lifetime.
\end{abstract}

\maketitle

\section{\label{Introduction}Introduction}

Magnons are excitations of ordered magnetic moments, which can perform logic operations and information transmission~\cite{magnonics1,magnonics3,manchon2019current,chumak2015magnon,demidov2017magnetization,bauer2012spin}. Nonlinear magnonics, which investigates nonlinear interactions among magnons and between magnons and other physical entities (phonons, photons, qubits, skyrmions, etc.), opens opportunities in magnon-based information processing beyond linear response regime. Parametric excitation of magnons is an efficient way to achieve nonlinear magnetization dynamics, which involves two typical configurations. 
One is the parallel pumping~\cite{parametrically,short,Fundamentals}, in which a microwave magnetic field of a frequency $\omega_d$ is parallel to the saturation magnetization, causing its photons to directly split into a pair of magnons with frequency $\omega_d/{2}$ and opposite wave vectors. 
The other is the perpendicular pumping~\cite{short,Fundamentals}, in which the microwave magnetic field perpendicular to the static magnetic field first excites the uniform ferromagnetic resonance (FMR) mode $(k=0)$; when the amplitude of the FMR mode exceeds a threshold, its energy is transferred to spin waves with $k\neq 0$, triggering the Suhl instability~\citep{Anderson,Suhl,Derivation}. 
In the first-order Suhl process, the FMR mode excites parametrically a pair of magnons with opposite wave vectors $\pm k$ and frequency $\omega_k=\omega_d/{2}$ through the three-magnon (dipolar) interactions~\citep{Modified,Saturation,Ultrashort,Three,Controlled}; in the second-order Suhl process, the FMR mode parametrically excites a pair of magnons with $\omega_k=\omega_d$ and opposite wave vectors $\pm k$ via the four-magnon (exchange) interactions~\citep{Direct,Compound}. Besides, the Kerr nonlinearity~\citep{induced,Dissipative,Magnetic,Long} between magnons due to the magnetocrystalline anisotropy~\cite{Magnetization} and cross-Kerr effects~\citep{Mechanical,Observation}  can also lead to bistability~\citep{Mechanical,Bistability,Theory,Explosive} and multi-stability~\citep{Long,Folding}.
Thereby, by utilizing various instability processes, magnons with selected frequencies and wave vectors can be pumped.

Recently, it was experimentally observed the anticrossing or mode-splitting phenomenon of the Walker spin-wave modes under strong microwave driving, which was phenomenologically explained through the strong coupling between a ``pump-induced magnon mode" and the Walker mode of the ferrimagnet~\cite{PIS}. The observed pumping-induced level repulsion of the magnon frequencies was later attributed to the oscillations between splitting and confluence in a three-magnon scattering process~\cite{Pumpinduced}. Arfini \textit{et al.}  constructed a theoretical framework of the magnonic three-wave mixing Hamiltonian to demonstrate that, under strong pumping at the FMR frequency, spectral splitting occurs as the driving amplitude increases~\cite{Magnon}.  These studies have only focused on mode splitting and anti-crossing phenomena, leaving the properties and dynamics of the pump-induced nonlinear modes unknown. This is because it is experimentally challenging to sensitively detect these pump-induced nonlinear modes since they do not couple with the probe microwaves. As one marked exception, Makiuchi \textit{et al.} found the lifetime of valley magnon pairs induced by the nonlinear magnetization dynamics is ultra-long using the transport method in magnetic film~\cite{long_lifetime}, a technique not applicable to the magnetic sphere.

\begin{figure}[htp!]
    \centering
    \includegraphics[width=1\linewidth]{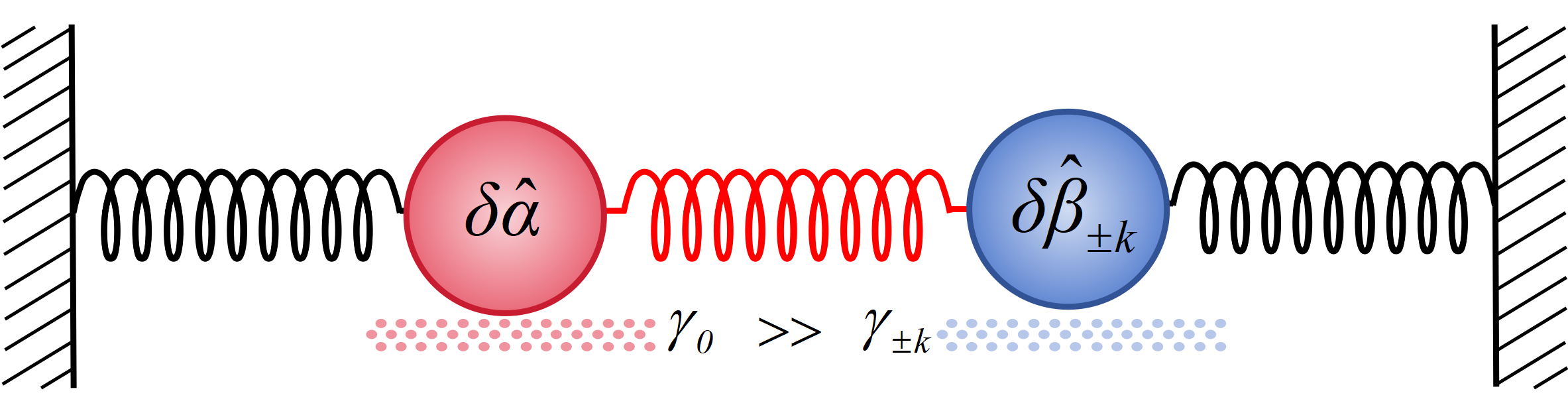}
    \caption{Coupled harmonic oscillators with different dampings. The red ball represents the fluctuation $\delta \hat{\alpha}$ of the Kittel magnon, and the blue ball represents fluctuations $\delta \hat{\beta}_{\pm k}$ of magnon pairs with wave vector $\pm k$. They hold very different damping $\gamma_0\gg \gamma_{\pm k}$.}
    \label{Fig.1}
\end{figure}

In this work, we probe the broadband nonlinear magnetization dynamics of a yttrium iron garnet (YIG) magnetic sphere under a strong microwave drive using microwave spectroscopy, focusing on the non-resonant pump regime, free of mode splitting, to extract the unique properties of the pump-induced magnon modes and their back action on FMR. \textit{Unexpectedly}, we find that when the driven amplitude is high and the pump frequency $\omega_d$ differs from the FMR frequency, the microwave transmission spectrum $|S_{21}|$ (dB) exhibits a sharp and asymmetric line shape, or a \textit{Fano resonance}~\cite{1961,nano,Topological,Controllable,2006}, a phenomenon usually existing due to the interference of discrete excited states and the continuum, thereby minimizing the absorption profile. This phenomena only appears in the nonlinear regime, and due to its steep and sharp asymmetric line shape~\cite{1961}, it differs entirely from the symmetric resonance phenomenon described by the Lorentz formula. We construct a scattering theory of photons and explain this phenomenon by involving the three-magnon interaction between the Kittel magnon and magnon pairs with opposite wave vectors of frequency $\omega_d/2$.  The driven steady-state amplitudes of these modes mediate an interaction between the fluctuation $\delta \hat{\alpha}$ of the Kittel magnon mode and the fluctuations $\delta \hat{\beta}_{\pm k}$ of the magnon pairs, as illustrated in Fig.~\ref{Fig.1}, which, according to our scattering theory, can be directly detected by the microwave transmission. We find that such Fano resonance occurs only when the damping coefficient $\gamma_{\pm k}$ of the fluctuation $\delta \hat{\beta}_{\pm k}$ is significantly smaller than the damping coefficient $\gamma_{0}$ of $\delta \hat{\alpha}$. This provides evidence that long-lifetime magnon pairs can be generated in the nonlinear magnetization dynamics. Our study provides a foundation for using the conventional microwave-spectroscopy approach to detect the back-action of other magnons on FMR and spin-wave resonance.

This article is organized as follows. In Sec.~\ref{Experiment}, we present the measured microwave transmission that shows the effects of pump frequency and pump power on the nonlinear magnetization dynamics of the YIG sphere.
In Sec .~\ref {theory}, we model the nonlinear magnetization dynamics by a quantum formalism involving the three-magnon interaction process and use the Lippmann-Schwinger formalism to derive the photon scattering matrix. In Sec.~\ref {calculation}, we compare the theoretical calculations with the measurements to explain the observed Fano resonance phenomenon. We summarize our results and give an outlook in Sec.~\ref{summary}.

\section{\label{Experiment}Experiment}

\begin{figure}[htp!]
\centering
\includegraphics[width=0.48\textwidth,trim=0.0cm 1cm 0cm 0.0cm\linewidth]{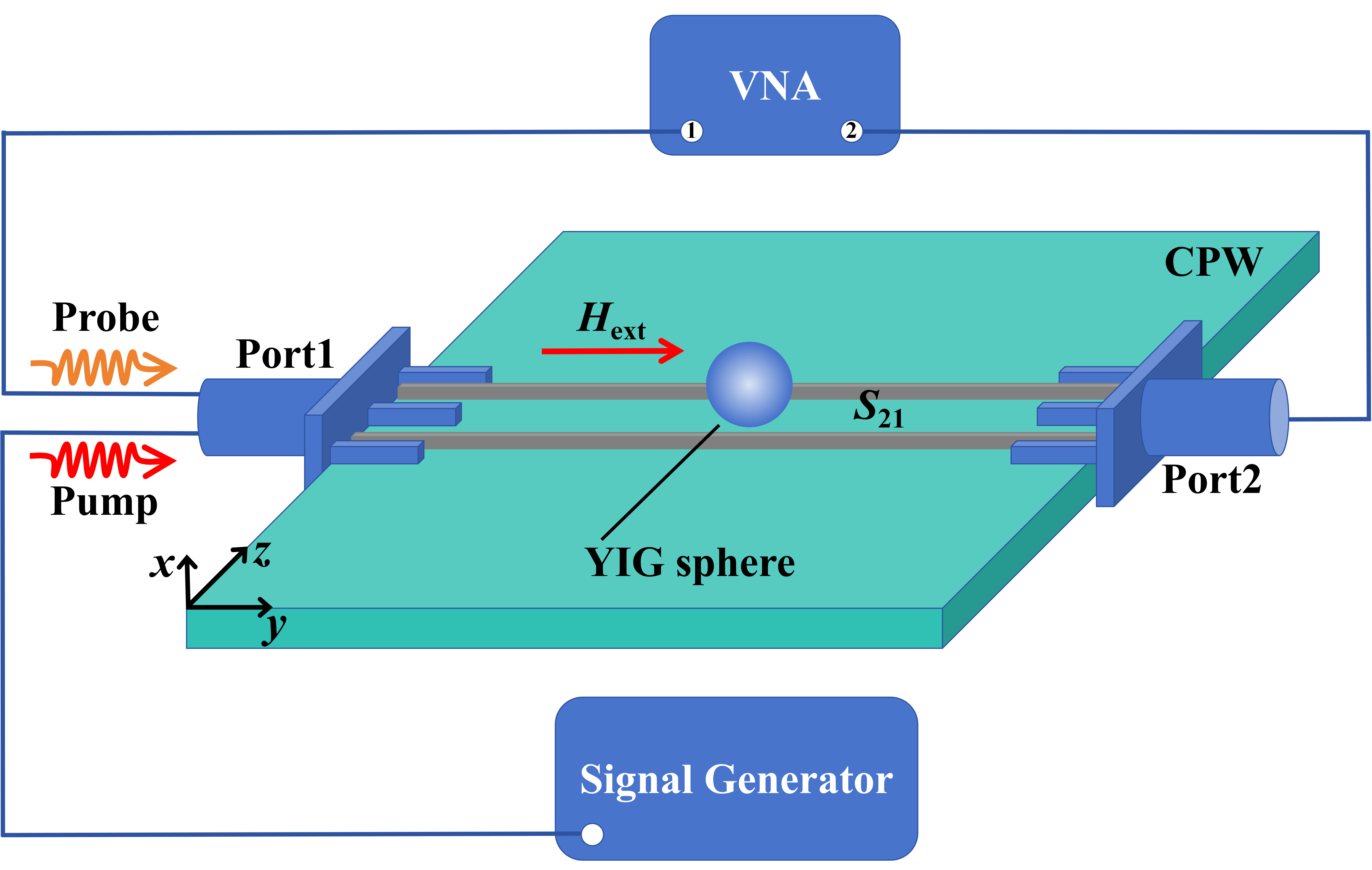}
\caption{Experimental configuration. A YIG sphere is loaded on top of the coplanar waveguide (CPW), biased by a magnetic field $H_{\rm ext }\hat{\bf y}$ along the central strip. The ``pump" and ``probe" microwave signals are generated, respectively, by the signal generator and vector network analyzer (VNA), which interact with the YIG sphere. The microwave transmission $S_{21}$ from Port ``1" to Port ``2" in the probe microwaves detects the dynamics of the magnetization when driven by the pump microwaves.}
\label{Fig.2}
\end{figure}

In the experiment, the coplanar waveguide (CPW) is manufactured on a copper substrate of lateral size 25~mm$\times$25~mm and thickness 0.03~mm, with a standard impedance of $50~\Omega$.
The CPW features a central strip width of 0.8~mm. We place a YIG sphere with a diameter of $1$~mm at the midpoint of the top surface of the central strip in the CPW to perform the measurements. A schematic illustration of the experimental measurement setup is illustrated in Fig.~\ref{Fig.2}. We position the CPW containing the YIG sphere horizontally at the center of the electromagnet, such that the direction of the external magnetic field $H_{\rm ext}\hat{\bf y}$ lies within the CPW plane and aligns with the central strip (using this as the $\hat{\bf y}$-axis to establish the coordinate system). We measure the microwave transmission $S_{21}$ from one terminal (Port ``1") to the other (Port ``2") of the CPW by using a vector network analyzer (VNA). Port ``1" of the VNA and a microwave signal generator are both connected via coaxial cables to Port ``1" of the CPW via a SubMiniature version A (SMA) connector (Fig.~\ref{Fig.2}). Port ``2" of the VNA is connected via a coaxial cable to Port ``2" of the CPW (Fig.~\ref{Fig.2}).

We set the power of the VNA to $-25$~dBm to minimize the additional impact of the microwave signal on the nonlinearity, ensuring it functions as a probe signal. Accordingly, we refer to the low-power microwave generated by the VNA as the ``probe" microwave, which scans a broadband frequency range $\omega_p$; whereas the microwave generated by the signal generator is referred to as the ``pump" microwave, whose frequency is restricted to a single frequency $\omega_d$ in one measurement. The power of the ``pump" microwave from the signal generator is $P_d$. During the experiment, we tune the VNA to a narrow intermediate frequency bandwidth (IFBW) of $1$~kHz; such narrowband filtering ensures that it only affects the pump frequency $\omega_d\pm 0.5$~kHz~\cite{microwaves}. On the other hand, the pump signal falls outside the receiver's passband and is effectively suppressed by the VNA's intermediate frequency filter.
 To investigate the nonlinear dynamics of the magnetization, we use the probe microwaves to measure the response of the magnetic sphere to a pump microwave of different frequencies and powers.

We first fix the external magnetic field at $\mu_{\rm 0} H_{\rm ext}=92.265$~mT, which saturates the sphere magnetization, and measure the microwave transmission spectrum $S_{21}$ of the system using the VNA. Due to the presence of the Walker modes in the YIG sphere, we observe several resonant dips~\cite{Magnetostatic,Ferrimagneticresonance,resonant2}.
We select the most significant dip at $2.780$~GHz as the subject of our experimental study. Since this dip agrees with the FMR formula of the magnetic sphere $\omega/2\pi=(\mu_0 \gamma H_{\rm ext}+\delta\omega)/2\pi$, where $\mu_0$ is the permeability of vacuum, $\gamma$ is the gyromagnetic ratio, and $\delta\omega$ is the inhomogeneous broadening, we identify it to be the FMR, i.e., the ferromagnetic resonance frequency $\omega_0/2\pi=2.780$~GHz.

Next, we fix the pump power at $P_{d}=-5$~dBm and change the pump frequency $\omega_d/2\pi$ from $2.7650$ to $2.7800$~GHz, i.e., the pump frequency $\omega_d<\omega_0$ approaches the FMR frequency. As shown in Fig.~\ref{Fig.3}(a),  when the pump frequency is far away from the FMR frequency, we find a tiny unstable signal as the frequency $\omega_d/2\pi$ varies from $2.7650$ to $2.7660$~GHz, which has little effect on the FMR absorption spectra. 

\begin{widetext}
\begin{center}
\begin{figure}
    \centering
    \includegraphics[width=0.9\linewidth]{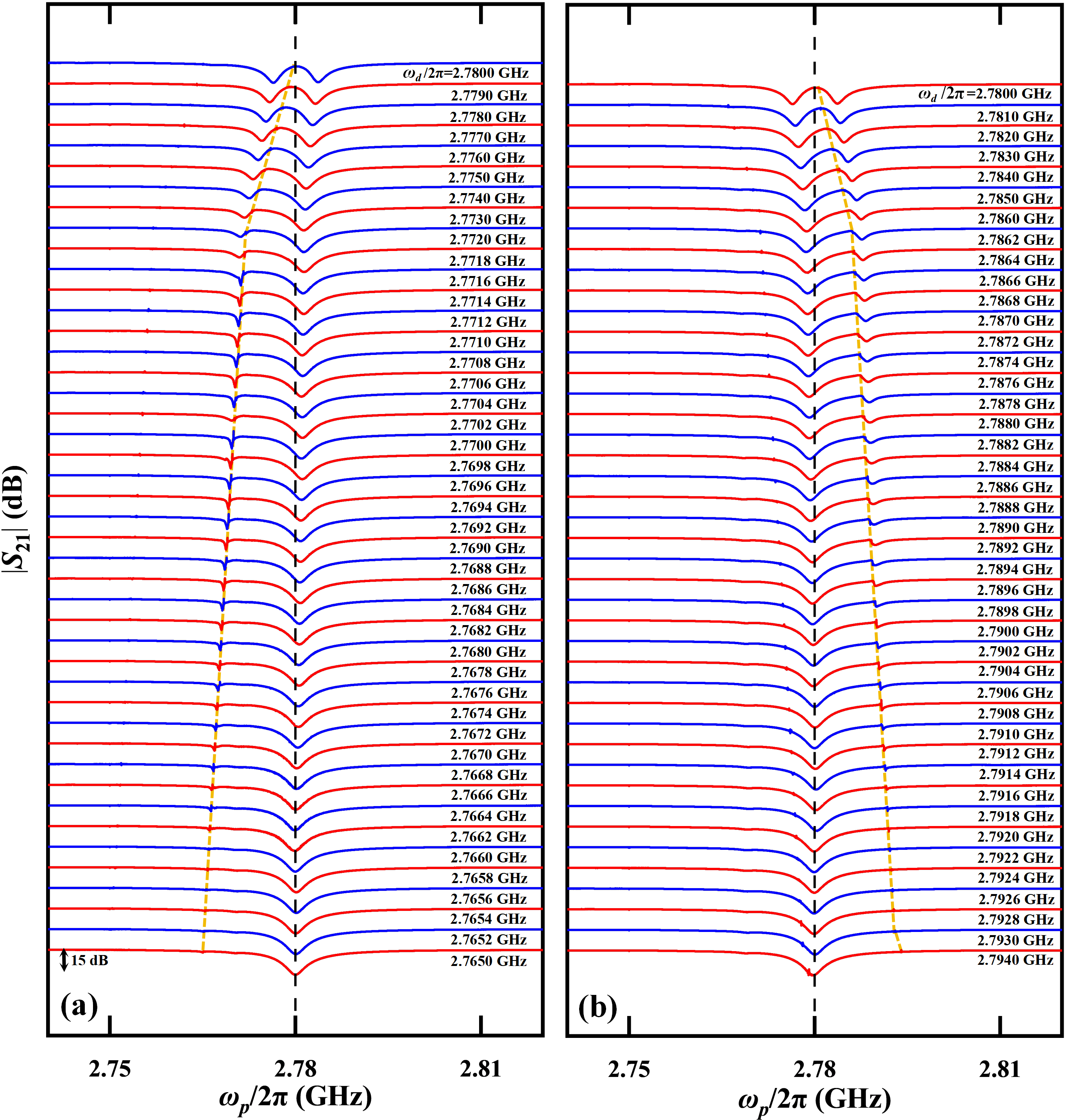}
    \caption{Measured microwave transmission spectra at a constant pump power $P_d=-5$~dBm and different pump frequencies. The black dashed curves indicate the FMR frequency $\omega_0/2\pi = 2.7800$ GHz, while the orange dashed curves represent the pump frequency $\omega_d/2\pi$. (a) When $\omega_d<\omega_0$, we change the pump frequency $\omega_d/2\pi$ from $2.7650$ to $2.7800$~GHz, i.e., the pump frequency approaches the FMR frequency. Unstable signals appear when $\omega_d/2\pi$ varies from $2.7650$ to $2.7660$~GHz; Fano resonance occurs from $2.7662$ to $2.7700$~GHz; the pump-induced mode splitting is observed from $2.7702$ to $2.7800$~GHz. 
    (b) When $\omega_d>\omega_0$, we change the pump frequency $\omega_d/2\pi$ from $2.7940$ to $2.7800$~GHz, i.e., away from the FMR frequency. Within this range, the FMR absorption spectrum is unaffected when the pump frequency is between $2.7940$ and $2.7920$~GHz; Fano resonance emerges when $\omega_d/2\pi$ exceeds $2.7918$~GHz; as the pump frequency increases, the Fano resonance gradually disappears while the modes split.}
    \label{Fig.3}
\end{figure}
\end{center}
\end{widetext}
Increasing the pump frequency to $\omega_d/2\pi=2.7662$~GHz, we observe a stable signal with an asymmetric spectral shape. 
This signal shows a sharp change between a dip and a peak, consistent with the shape of the Fano resonance~\citep{Topological,Controllable,1961}.
As the pump frequency shifts from $2.7662$ to $2.7700$~GHz, the characteristics of the Fano resonance become increasingly apparent. We continue to increase the pump frequency to $2.7702$~GHz and find a dip appears below the pump frequency, which agrees with the reported ``pump-induced splitting" by Rao \textit{et al.}~\cite{PIS}. The magnetic sphere, however, is saturated in our sample, as we confirm via measuring the unchanged FMR frequency in different hysteresis loops of the magnetization. At the same time, we observe that as the pump frequency gradually approaches the FMR frequency, the depth of the dip in the microwave transmission near the pump frequency $\omega_d/2\pi$ increases. At the FMR frequency $\omega_d/2\pi=2.7800$~GHz, the single resonance dip dramatically splits into two dips with equal intensities, in agreement with previously reported pump-induced splitting observations~\cite{PIS,Pumpinduced,Magnon}.

Further, as shown in Fig.~\ref{Fig.3}(b), we modulate the pump frequency to approach the FMR frequency when $\omega_d>\omega_0$. When the pump is far from the FMR, we again observe unstable signals at the pump frequency, without Fano resonance or pump-induced splitting phenomena.
When the pump frequency $\omega_d/2\pi=2.7918$~GHz, we also observe the phenomenon of Fano resonance. However, comparing Fig.~\ref{Fig.3}(a) and (b), we find that the two Fano resonances are \textit{completely opposite}: 
We observe that the signal in Fig.~\ref{Fig.3}(a) first exhibits a dip followed by a peak, while the spectrum in Fig.~\ref{Fig.3}(b) shows a peak first and then descends into a dip. As we continue to decrease the pump frequency, we observe that the Fano-resonance phenomenon first becomes more pronounced, then evolves into the mode splitting phenomenon.

 We summarize the above data and plot the systematic dependence of microwave transmission on the pump and probe frequencies in Fig.~\ref{Fig.4}. Region I in Fig.~\ref{Fig.4}(a) corresponds to the unstable signal that appears when $\omega_d/2\pi$ varies from $2.7650$ to $2.7660$~GHz in Fig.~\ref{Fig.3}. As already addressed in Fig.~\ref{Fig.3}, we observe Fano resonance occurring in region II of Fig.~\ref{Fig.4}(a), with $\omega_d/2\pi$ ranging from $2.7662$ to $2.7700$~GHz. To more clearly address the characteristics of the Fano resonance, Fig.~\ref{Fig.4}(b) enlarges the map of microwave transmission extracted from region II in Fig.~\ref{Fig.4}(a). As shown in Fig.~\ref{Fig.4}(b), when $\omega_d<\omega_0$, a sharp change from a dip to a peak occurs with $\omega_p$, exhibiting characteristics consistent with the line shape of Fano resonance. As we continue to increase $\omega_d/2\pi$ to $2.7800$~GHz, as shown in region III of Fig.~\ref{Fig.4}(a), we observe the pump-induced splitting, with the dip near the pump frequency deepening while the dip near the FMR frequency becomes shallower. At the same time, two dips of equal intensity appear when $\omega_d=\omega_0$.

\begin{widetext}
\begin{center}
\begin{figure}[htp!]
    \centering
    \includegraphics[width=1\linewidth]{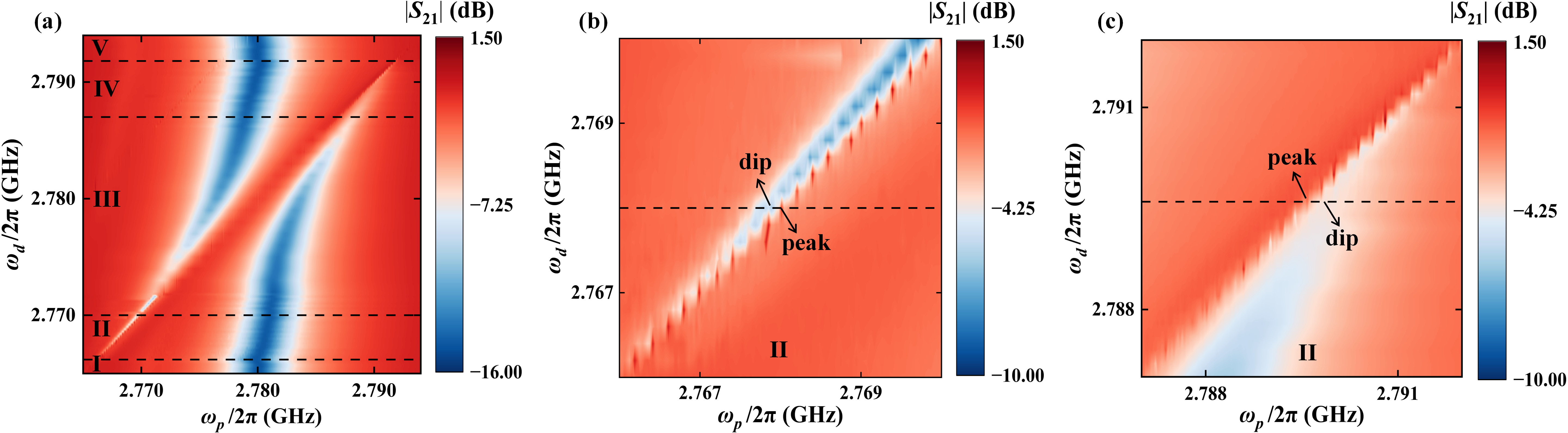}
    \caption{Map of microwave transmissions $|S_{21}|$ as a function of different pump $\omega_d$ and probe $\omega_p$ frequencies at a constant pump power $P_d=-5$~dBm. (a) is plotted with the pump frequency $\omega_d/2\pi$ varying from $2.7650$ to $2.7940$~GHz, across the FMR frequency $\omega_{0}/2\pi=2.7800$~GHz. Unstable signals appear in region I when $\omega_d/2\pi$ varies from $2.7650$ to $2.7660$~GHz; region II is the Fano resonance, occurring from 2.7662 to 2.7700 GHz;  the pump-induced mode splitting is observed from $2.7702$ to $2.7870$~GHz in region III; Fano resonance emerges in region IV when $\omega_d/2\pi$ exceeds $2.7870$~GHz; the FMR absorption spectrum is unaffected in region V when the pump frequency is between $2.7920$ and $2.7940$~GHz. (b) is an enlarged view of the Fano resonance in region II of (a). (c) An enlarged view of the Fano resonance in region IV of (a).}
    \label{Fig.4}
\end{figure}
\end{center}
\end{widetext}

Furthermore, as the pump frequency increases beyond the mode-splitting regime III, Fano resonance begins to occur again, corresponding to region IV of Fig.~\ref{Fig.4}(a). To further study the specific characteristics of the Fano resonance, we plot the enlarged region IV in Fig.~\ref{Fig.4}(c). By comparing Fig.~\ref{Fig.4}(b) and (c), we clearly observe that Fano resonance exhibits two completely opposite line profiles: Fig.~\ref{Fig.4}(b) shows a dip followed by a peak, while Fig.~\ref{Fig.4}(c) displays a peak followed by a dip. Finally, in region V of Fig.~\ref{Fig.4}(a), we observe that the FMR frequency remains consistently stable, indicating that the FMR is unaffected when the pump frequency is far above the FMR frequency.

\begin{figure}[htp!]
    \centering
    \includegraphics[width=1\linewidth]{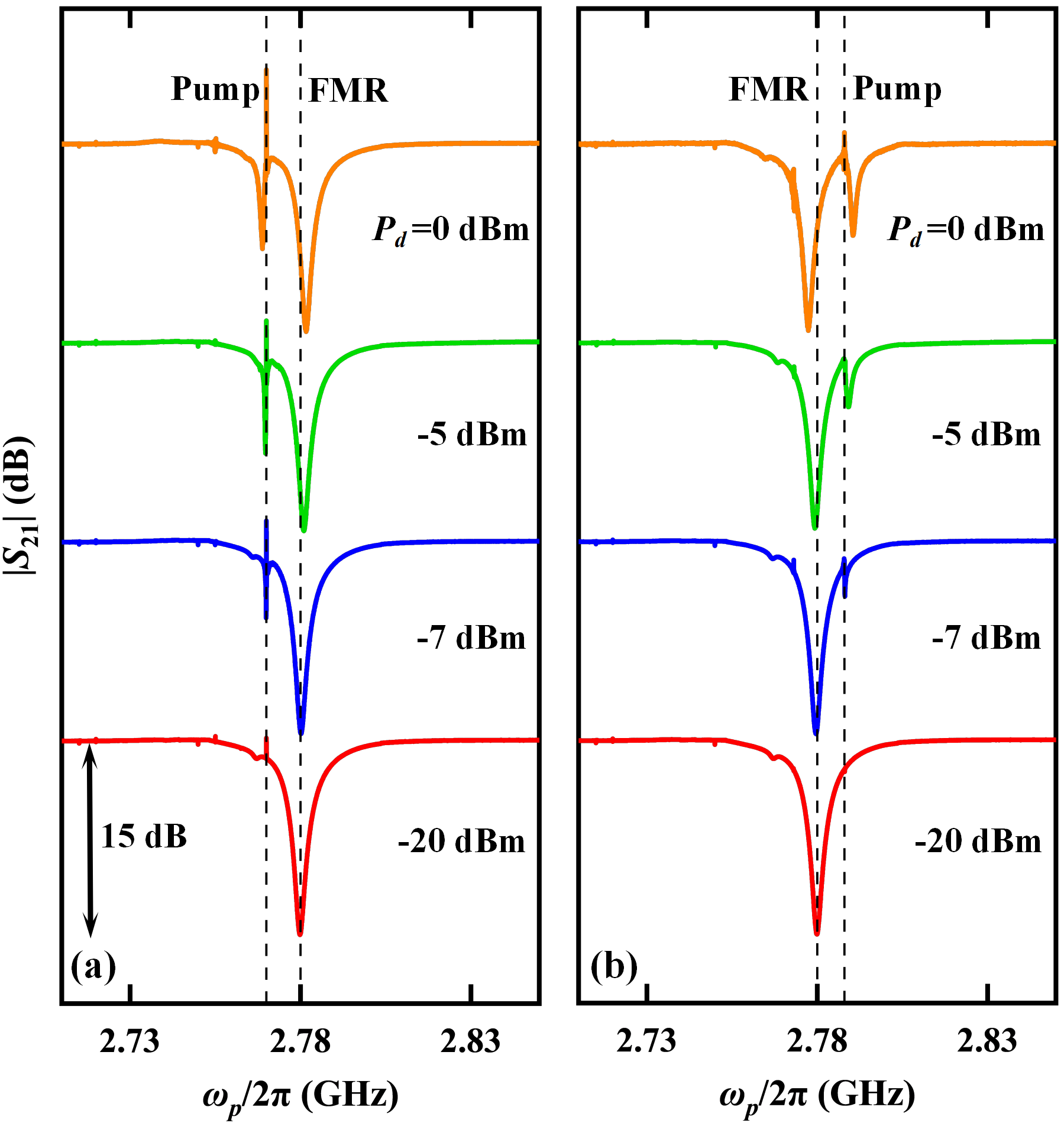}
    \caption{Measured microwave transmission spectra at different pump powers under a fixed pump frequency. (a) When the fixed pump frequency is set to $\omega_d/2\pi=2.770$~GHz (below the FMR frequency), the pump power $P_d=-20$~dBm causes a signal instability at $\omega_d/2\pi$; increasing the pump power to $P_d=-7$~dBm induces the Fano resonance that persists until $P_d=0$~dBm. (b) When the pump frequency is fixed at $\omega_d/2\pi=2.788$~GHz (above the FMR frequency), no significant phenomenon appears at $P_d=-20$~dBm; when $P_d=-7$~dBm, a Fano resonance with the shape opposite to that observed at $\omega_d/2\pi=2.770$~GHz emerges; as the power increases to $P_d=0$~dBm, the Fano resonance characteristics become more pronounced.}
    \label{Fig.5}
\end{figure}

We then turn to address the pump-power dependence of the nonlinear magnetization dynamics by fixing the pump frequencies at $\omega_d/2\pi=2.770$~GHz and $\omega_d/2\pi=2.788$~GHz, which are below and above the FMR frequency $\omega_0/2\pi=2.780$~GHz. When the power $P_d=-20$~dBm for the pump microwave of frequency $\omega_d/2\pi=2.770$~GHz, the dip of FMR does not change, but a fluctuating signal appears at the pump frequency $\omega_d/2\pi=2.770$~GHz, as shown in Fig.~\ref{Fig.5}(a).
When we increase the pump power to $P_d=-7$~dBm, the transmission spectrum $|S_{21}|$ exhibits a significant change, i.e., the emergence of a Fano resonance phenomenon characterized by a dip-then-peak structure. Increasing the pump power to $P_d=-5$~dBm and $0$~dBm, we find that the signal exhibits a more pronounced Fano resonance. At the same time, we observe that the FMR absorption spectrum shifts to a higher frequency.

When the pump frequency is fixed at $\omega_d/2\pi=2.788$~GHz, which is larger than the FMR frequency, we vary the pump power from $P_d=-20$~dBm to 0~dBm. When the pump power is low, Fig.~\ref{Fig.5}(b) shows the same phenomenon as Fig.~\ref{Fig.5}(a), i.e., an unstable signal appears that does not affect the FMR absorption spectrum. When the pump power is increased to $P_d=-7$~dBm and $0$~dBm, we observe Fano resonance at both pump powers. In contrast to Fig.~\ref{Fig.5}(a), the Fano resonance in the spectra of Fig.~\ref{Fig.5}(b) shows opposite symmetry, characterized by a peak followed by a dip, and the FMR absorption is shifted to a lower frequency.

In a short summary, when the pump frequency is far from the FMR frequency, we observe an unstable signal at the pump frequency in the microwave transmission, as shown in Fig.~\ref{Fig.3}. As the pump frequency approaches the FMR, we observe the appearance of a Fano resonance around the drive frequency (Fig.~\ref{Fig.3} and \ref{Fig.5}). When the pump frequency is close to the FMR frequency, we observe the phenomenon of pump-induced splitting. To analyze the generation of Fano resonances and pump-induced splitting and to explain their evolution with the pump frequency and pump power, we establish the following theoretical model and scattering theory of microwave photons.

\section{\label{theory}Scattering theory for detection of nonlinear magnon modes}

\subsection{Pump-induced magnon interaction}

\begin{figure}[htbp]
    \centering
    \includegraphics[width=0.87\linewidth]{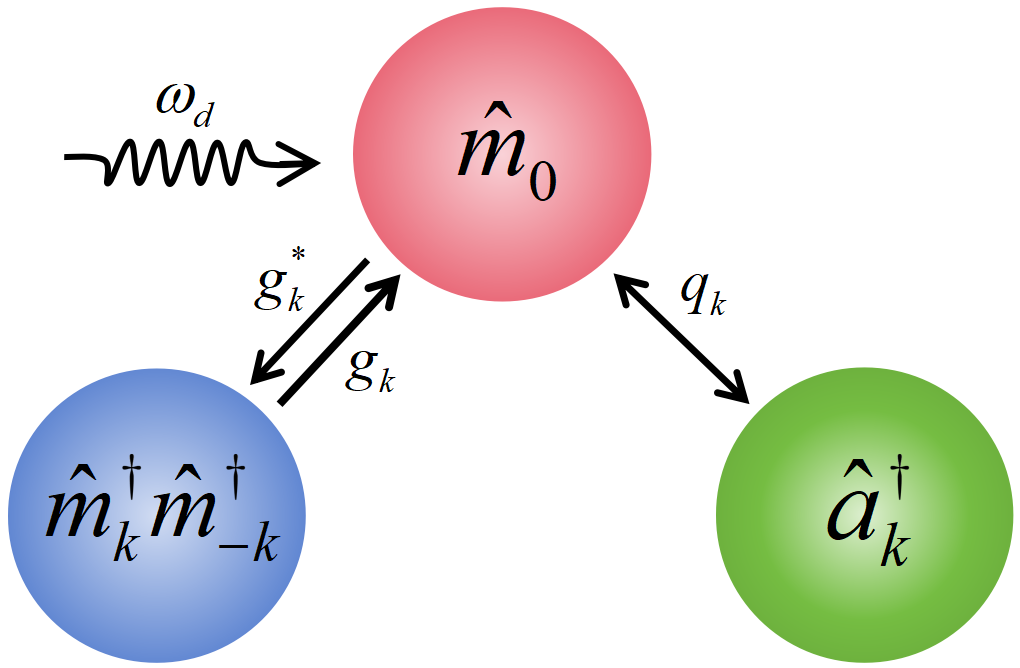}
    \caption{Three-magnon interaction in the magnetic sphere driven by a strong ``pump" microwave of frequency $\omega_d$. $\hat{m}_{0}$ and $\hat a_{k}$ represent, respectively, the Kittel magnon and the microwave photon, coupled via a coupling strength $q_k$. $\hat{m}_{k}$ and $\hat{m}_{-k}$ are a pair of magnons with opposite wave vectors, which couple to $\hat{m}_0$ with the coupling strengths $g_k$.}
    \label{Fig.6}
\end{figure}

To explain the experimental observation, we set up a model involving the three-magnon interaction~\citep{Modified,Saturation,Ultrashort,Three,Controlled}. As illustrated in Fig.~\ref{Fig.6}, we consider the Kittel magnon $\hat{m}_{0}$  of the FMR frequency $\omega_{0}$, which via the coupling constant $g_k$ couples \textit{nonlinearly} with two magnon modes with opposite wave vectors $\pm k$ of frequency $\omega_k$, one being $\hat{m}_{k}$ and the other being $\hat{m}_{-k}$. Physically, such three-magnon interaction originates from the magnetic dipole-dipole coupling~\citep{Modified,Saturation}, known in the first-order Suhl instability. The uniform magnetization precession originating from the ferromagnetic resonance generates the dipolar field antiparallel to the magnetization, which couples to the magnon pairs that carries zero net wave vector, thereby obeying the energy conservation $\hbar\omega_0 = \hbar\omega_k + \hbar\omega_{-k}$ and momentum conservation $0 = \bf{k} + (-\bf{k})$. Due to specific boundary conditions and demagnetization fields present in the YIG sphere, a dense distribution of magnetostatic modes appears below the FMR mode~\citep{Modes}. Such a rich spectrum of modes provides the phase space for the splitting of the FMR into two half-frequency magnons ($\omega_0/2$) via the dipolar interaction. The Kittel magnon $\hat{m}_0$ is under the coherent drive of the drive frequency $\omega_{d}$ with the drive strength $\Omega_{d}$ by the ``pump" microwaves, which is probed by the photon modes $\hat a_{k}$ of frequency $\Omega_{k}=\omega_{p}$ in the coplanar waveguide. 
According to the input-output formalism~\citep{Folding,origin,Chiral}, $\Omega_{d}=\sqrt{{P_{d} \gamma_{\rm ext}}/({2 \hbar \omega_0}})$ is related to the input microwave power $P_{d}$ and the external losses of the resonantly driven Kittel mode.

Such a model is described by the Hamiltonian 
\begin{align}
\label{Hamiltonian}
    \hat H/\hbar&=\omega_{0} \hat{m}_{0}^{\dagger} \hat{m}_{0}+\omega_{k} \hat{m}_{k}^{\dagger} \hat{m}_{k}+\omega_{-k} \hat{m}_{-k}^{\dagger} \hat{m}_{-k}\nonumber\\
    &+g_{k}^{*} \hat{m}_{0} \hat{m}_{k}^{\dagger} \hat{m}_{-k}^{\dagger}
    +g_{k} \hat{m}_{0}^{\dagger} \hat{m}_{k} \hat{m}_{-k}\nonumber\\
    &+i\left(\Omega_{d}^{*} \hat{m}_{0}e^{i\omega_{d}t}-\Omega_{d} \hat{m}_{0}^{\dagger}e^{-i\omega_{d}t}\right)\nonumber\\
&+\sum_k\Omega_k\hat{a}_k^{\dagger}\hat{a}_k+\sum_kq_k\left(\hat{a}_k^\dagger\hat{m}_0+\hat{a}_k\hat{m}^\dagger_0\right),
\end{align}
where $g_{k}$ is the coupling strength between the magnetostatic Kittel mode and the magnon pairs with wave number $\pm k$, $q_{k}$ is the coupling strength between the Kittel mode and the photon mode in the coplanar waveguide. In this minimal model, ${\bf k}$ is understood as the quantum number of one dominant magnon pair that couples to the Kittel mode, without involving the complication of dense degenerate magnetostatic modes of a YIG sphere.
According to the Heisenberg equation, we obtain the equation of motion of these operators 
\begin{align}
&\frac{d \hat{m}_{0}}{dt}=\left(-i\omega_{0}-\frac{\gamma_{0}}{2}\right)\hat{m}_{0}-ig_{k}\hat{m}_{k}\hat{m}_{-k}\nonumber \\
&~~~~~~-\Omega_{d}e^{-i \omega_{d} t}-i\sum_k q_k\hat{a}_k,\nonumber \\
&\frac{d \hat{m}_{k}}{dt}=\left(-i\omega_{k}-\frac{\gamma_{k}}{2}\right)\hat{m}_{k}-ig_{k}^{*} \hat{m}_{0}\hat{m}_{-k}^{\dagger},\nonumber \\
&\frac{ d \hat{m}_{-k}}{dt}=\left(-i\omega_{-k}-\frac{\gamma_{-k}}{2}\right)\hat{m}_{-k}-ig_{k}^{*} \hat{m}_{0} \hat{m}_{k}^{\dagger},\nonumber \\
&\frac{d \hat{a}_{k}}{dt}=-i q_k\hat{m}_0-i\Omega_k\hat{a}_k,
\label{equation_of_motion}
\end{align}
in which $\gamma_{0}$ accounts for the dissipation of the Kittel mode and $\gamma_{\pm k}$ is the dissipation of the magnons with opposite wave number $\pm k$.

To eliminate the fast oscillation in the pumping, we express $\hat{m}_{0}(t)=\hat\alpha(t) e^{-i \omega_{d} t}$, $\hat{m}_{k}(t)=\hat\beta _{k}(t)e^{-i{\omega_{d} t}/{2}}$, $\hat{m}_{-k}(t)=\hat\beta _{-k}(t)e^{-i{\omega_{d}t}/{2}}$, and $\hat{a}_{k}(t)=\hat c_{k}(t) e^{-i \omega_{d} t}$ in terms of oscillators $\{ \hat\alpha(t), \hat\beta _{k}(t), \hat\beta _{-k}(t), c_{k}(t)\}$. 
The constant drive creates the steady-state amplitudes, for which we construct a mean-field solution via setting $\hat{\alpha}=\langle\hat{\alpha}\rangle+\delta \hat{\alpha}$, $\hat{\beta}_{k}=\langle\hat{\beta}_{k}\rangle+\delta \hat{\beta}_{k}$, $\hat{\beta}_{-k}=\langle\hat{\beta}_{-k}\rangle+\delta \hat{\beta}_{-k}$, and  $\hat{c}_{k}=\langle\hat{c}_{k}\rangle+\delta \hat{c}_{k}$, where $\langle\hat{\alpha}\rangle$, $\langle\hat{\beta}_{k}\rangle$, $\langle\hat{\beta}_{-k}\rangle$, and $\langle\hat{c}_{k}\rangle$ are the steady-state solution independent of time, while $\delta \hat{\alpha}$, $\delta \hat{\beta}_{k}$, $\delta \hat{\beta}_{-k}$, and $\delta \hat{c}_{k}$ are the time-dependent fluctuations.  
Substitution into Eq.~\eqref{equation_of_motion} leads to the equation of motion for the mean-field solution: 
\begin{align}
\label{eq3}
&\frac{ d \langle\hat\alpha\rangle}{dt}=-\left[i(\omega_{0}-\omega_{d})+\frac{\gamma_{0}}{2}\right]\langle\hat\alpha\rangle-ig_{k}\langle\hat\beta _{k}\rangle\langle\hat\beta _{-k}\rangle\nonumber\\
&~~~~~~~~-\Omega_{d}-i\sum_kq_{k}\langle\hat c_{k}\rangle=0,\nonumber\\
&\frac{d \langle\hat\beta _{k}\rangle}{dt}=-\left[i\left(\omega_{k}-\frac{\omega_{d}}{2}\right)+\frac{\gamma_{k}}{2}\right]\langle\hat\beta _{k}\rangle-ig_{k}^{*}\langle\hat\alpha\rangle \langle\hat\beta _{-k}^{\dagger}\rangle=0,\nonumber\\
&\frac{d \langle\hat\beta _{-k}\rangle}{d t}=-\left[i\left(\omega_{-k}-\frac{\omega_{d}}{2}\right)+\frac{\gamma_{-k}}{2}\right]\langle\hat\beta _{-k}\rangle-ig_{k}^{*}\langle\hat\alpha\rangle \langle\hat\beta _{k}^{\dagger}\rangle=0,\nonumber\\
&\frac{d \langle\hat c_{k}\rangle}{dt}=-i(\Omega_k-\omega_{d})\langle\hat c_{k}\rangle-iq_{k}\langle\hat\alpha\rangle=0.
\end{align}
Considering the situation without the probe, i.e., disregarding $\langle\hat{c}_k\rangle$, we obtain  the steady-state solutions 
\begin{align}
\label{amplitudes}
&\langle\hat\alpha\rangle=-\frac{ig_{k}\langle\hat\beta _{k}\rangle\langle\hat\beta _{-k}\rangle+\Omega_{d}}{i(\omega_{0}-\omega_{d})+{\gamma_{0}}/{2}},\nonumber\\
&\langle\hat\beta _{k}\rangle=-\frac{ig_{k}^{*}\langle\hat\alpha \rangle\langle\hat\beta _{-k}^{\dagger}\rangle}{i(\omega_{k}-{\omega_{d}}/{2})+{\gamma_{k}}/{2}},\nonumber\\
&\langle\hat\beta _{-k}\rangle=-\frac{ig_{k}^{*}\langle\hat\alpha\rangle \langle\hat\beta _{k}^{\dagger}\rangle}{i(\omega_{-k}-{\omega_{d}}/{2})+{\gamma_{-k}}/{2}}.
\end{align}

Based on these solutions, we find the relations 
\begin{subequations}
    \begin{align}
\frac{\langle\hat\beta _{k}\rangle}{\langle\hat\beta _{-k}^{\dagger}\rangle}&=-\frac{ig_{k}^{*}\langle\hat\alpha \rangle}{i(\omega_{k}-{\omega_{d}}/{2})+{\gamma_{k}}/{2}},\\
\frac{\langle\hat\beta _{-k}\rangle}{\langle\hat\beta _{k}^{\dagger}\rangle}&=-\frac{ig_{k}^{*}\langle\hat\alpha\rangle }{i(\omega_{-k}-{\omega_{d}}/{2})+{\gamma_{-k}}/{2}}.
\end{align}
\end{subequations}
Accordingly, we obtain $|\langle\hat\beta _{k}\rangle|^{2}=|\langle\hat\beta _{-k}\rangle|^{2}$. 
Further, when we express the steady-state solutions in the
form of $\langle\hat\beta _{k}\rangle=\beta e^{i\phi_{k}}$, $\langle\hat\beta_{-k}\rangle=\beta e^{i\phi_{-k}}$, and the coupling in the form of $g _{k}=| g_{k} |e^{i\phi_{g}}$, the total phase can be defined as $ \phi=\phi_{k}+\phi_{-k}+\phi_{g}$~\cite{Pumpinduced}. Hence, the steady-state equations are rewritten as 
\begin{subequations}
\begin{align}
&\left[i (\omega_{0}-\omega_{d})+\gamma_{0} / 2\right]\left\langle\hat{\alpha}\right\rangle+\Omega_{d}+i\left|g_{k}\right| \beta^{2} e^{i \phi}=0 ,\\
&\left[i (\omega_{k}-\frac{\omega_{d}}{2})+\gamma_{k} / 2\right] \beta e^{i \phi}+i\left|g_{k}\right|\left\langle\hat{\alpha}\right\rangle \beta=0.
\end{align}
\end{subequations}
Eliminating $\langle\hat{\alpha}\rangle$ leads to the equation for $\beta$:
\begin{align}
&\left[i (\omega_{k}-{\omega_{d}}/{2})+\gamma_{k} / 2\right]\left[i (\omega_{0}-\omega_{d})+\gamma_{0} / 2\right] e^{i \phi}\nonumber\\
&=i |g _{k}|\Omega_{d} -|g_{k}|^{2}\beta^{2}e^{i \phi}.
\label{equation_beta}
\end{align}

An analytical solution of Eq.~\eqref{equation_beta} can be obtained when the drive frequency coincides with the FMR resonance, i.e., $ \omega_{d}=\omega_{0}$. At the resonance conditions $\omega_d = \omega_0$ and $\omega_k = \omega_{-k} = \omega_d/2$, Eq.~\eqref{equation_beta} leads to $\left({\gamma_k \gamma_0}/{4} + |g_k|^2 \beta^2 \right) e^{i\phi} = i|g_k|\Omega_d$, which indicates that the phase $\phi = \pi/2$ and $\beta$ is real. In this case, the steady-state value of the magnon pairs can be solved explicitly, giving~\cite{Magnon,Pumpinduced}
\begin{align}
 \beta=\sqrt{\frac{4 |g_{k} |\Omega_{d}-\gamma_{0} \gamma_{k}}{4 |g_{k}|^{2}}}.
 \label{beta}
\end{align}
However, when $\omega_d$ is away from $\omega_0$, we have to find a numerical solution. We obtain the driven amplitudes $\langle\alpha\rangle$ and $\langle\hat{\beta}_{\pm k}\rangle$ by numerically solving the full set of coupled algebraic equations Eq.~\eqref{eq3} directly. The resulting amplitudes mediate an interaction between the fluctuations $\delta \hat{\alpha}$, $\delta \hat{\beta}_{k}$, $\delta \hat{\beta}_{-k}$, and $\delta \hat{c}_{k} $, which after disregarding the terms of higher orders obey
\begin{align}
\label{delta}
&\frac{d \delta \hat{\alpha}}{dt}=-\left[i(\omega_{0}-\omega_{d})+\frac{\gamma_{0}}{2}\right]\delta \hat{\alpha}\nonumber\\
&~~~~~~~~-ig_{k}\left(\langle\hat{\beta}_{k}\rangle \delta \hat{\beta}_{-k}+\delta \hat{\beta}_{k}\langle\hat{\beta}_{-k}\rangle\right)-i\sum_kq_{k}\delta \hat{c}_{k},\nonumber\\
&\frac{d \delta \hat{\beta}_{k}}{d t}=-\left[i\left(\omega_{k}-\frac{\omega_{d}}{2}\right)+\frac{\gamma_{k}}{2}\right]\delta \hat{\beta}_{k}\nonumber\\
&~~~~~~~~-ig^{*}_{k}\left(\left\langle\hat{\alpha}\right\rangle \delta \hat{\beta}^{\dagger}_{-k}+\delta \hat{\alpha}\langle\hat{\beta}^{\dagger}_{-k}\rangle\right),\nonumber\\
&\frac{d \delta \hat{\beta}_{-k}}{d t}=-\left[i\left(\omega_{-k}-\frac{\omega_{d}}{2}\right)+\frac{\gamma_{-k}}{2}\right]\delta \hat{\beta}_{-k}\nonumber\\
&~~~~~~~~~-ig^{*}_{k}\left(\left\langle\hat{\alpha}\right\rangle \delta \hat{\beta}^{\dagger}_{k}+\delta \hat{\alpha}\langle\hat{\beta}^{\dagger}_{k}\rangle\right),\nonumber\\
&\frac{d \delta \hat{c}_k}{dt}=-i(\Omega_k-\omega_{d})\delta \hat c_{k}-iq_{k}\delta \hat{\alpha}.
\end{align} 
According to Eq.~\eqref{delta}, we construct the effective Hamiltonian for the fluctuations as 
\begin{align}
    &\hat H_{\rm eff}/\hbar=\left(\omega_{0}-\omega_{d}\right) \delta\hat{\alpha}^{\dagger} \delta\hat{\alpha}+\left(\omega_{-k}-\frac{\omega_{d}}{2}\right) \delta\hat{\beta}_{-k}^{\dagger} \delta\hat{\beta}_{-k}\nonumber\\
    &+\left(\omega_{k}-\frac{\omega_{d}}{2}\right) \delta\hat{\beta}_{k}^{\dagger} \delta\hat{\beta}_{k}+g_{k}\langle\hat{\beta}_{k}\rangle \delta \hat{\beta}_{-k}\delta\hat{\alpha}^{\dagger}\nonumber\\
&+g^{*}_{k}\langle\hat{\beta}^{\dagger}_{k}\rangle\delta \hat{\beta}^{\dagger}_{-k}\delta\hat{\alpha}+g_{k}\langle\hat{\beta}_{-k}\rangle \delta\hat{\beta}_{k}\delta\hat{\alpha}^{\dagger}\nonumber\\&+g^{*}_{k}\langle\hat{\beta}^{\dagger}_{-k}\rangle \delta \hat{\beta}^{\dagger}_{k}\delta\hat{\alpha}+g^{*}_{k}\left\langle\hat{\alpha}\right\rangle \delta \hat{\beta}^{\dagger}_{-k}\delta\hat{\beta}^{\dagger}_{k}+g_{k}\left\langle\hat{\alpha}^{\dagger}\right\rangle \delta \hat{\beta}_{-k}\delta\hat{\beta}_{k}\nonumber\\
    &+\sum_k(\Omega_k-\omega_{d})\delta \hat c_{k}^{\dagger}\delta \hat c_{k}+\sum_kq_{k}(\delta \hat{c}_{k}\delta \hat{\alpha}^{\dagger}+\delta \hat{c}^{\dagger}_{k}\delta \hat{\alpha}).
\end{align} 
It shows several features. On one hand, $\delta \hat{\alpha}$ couples with $\delta\hat{\beta}_k$ ($\delta\hat{\beta}_{-k}$) via the coupling constant $g_k\langle \hat{\beta}_{-k}\rangle$ ($g_k\langle \hat{\beta}_{k}\rangle$); $\delta \hat{\beta}_k$ couples with $\delta \hat{\beta}^{\dagger}_{-k}$ via the coupling constant $g_k^*\langle\alpha\rangle$. On the other hand, the fluctuation of VNA microwaves $\delta \hat{c}_k$ couples with $\delta\hat{\alpha}$, which can thereby directly detect the fluctuation of ferromagnetic resonance, but can only detect $\delta\hat{\beta}_{\pm k}$ indirectly.

\subsection{Microwave scattering matrix}

As addressed above, the fluctuation of the VNA microwaves $\delta \hat{c}_k$ couples directly with the fluctuation $\delta\hat{\alpha}$ and thereby detects its dynamics, which reflects the back action and properties of $\delta\hat{\beta}_{\pm k}$ due to their interaction with $\delta\hat{\alpha}$ mediated by the driven amplitudes $\langle\alpha\rangle$ and $\langle\hat{\beta}_{\pm k}\rangle$. To calculate the scattering matrix of the probe microwave photon, we decompose the effective Hamiltonian $\hat{H}_{\rm eff}$ into an uncoupled free Hamiltonian $\hat H_{0}$ and interaction Hamiltonian $\hat H_{\rm int}$:
\begin{align}
    &\hat H_{0}/\hbar=\left(\omega_{0}-\omega_{d}\right) |{\delta\alpha}\rangle \langle{\delta\alpha}|+\left(\omega_{k}-\frac{\omega_{d}}{2}\right) |{\delta\beta}_{k}\rangle \langle{\delta\beta}_{k}|\nonumber\\
    &+\left(\omega_{-k}-\frac{\omega_{d}}{2}\right)|\delta{\beta}_{-k}\rangle  \langle{\delta\beta}_{-k}|+\sum_k(\Omega_k-\omega_{d})|\delta c_{k}\rangle \langle\delta c_{k}|,\nonumber\\
    &\hat H_{\rm int}/\hbar=g_{k}\langle\hat{\beta}_{k}\rangle| {\delta\alpha}\rangle \langle{\delta\beta}_{-k}|+g^{*}_{k}\langle\hat{\beta}^{\dagger}_{k}\rangle|{\delta\beta}_{-k}\rangle\langle{\delta\alpha}|\nonumber\\
    &+g_{k}\langle\hat{\beta}_{-k}\rangle |{\delta\alpha}\rangle\langle{\delta\beta}_{k}|+g^{*}_{k}\langle\hat{\beta}^{\dagger}_{-k}\rangle |\delta{\beta}_{k}\rangle\langle\delta{\alpha}|\nonumber\\
    &+g^{*}_{k}\left\langle\hat{\alpha}\right\rangle (|\delta{\beta}_{-k}\rangle\otimes|\delta{\beta}_{k}\rangle\langle 0|)\nonumber\\
&+g_{k}\left\langle\hat{\alpha}^{\dagger}\right\rangle (|0\rangle \langle\delta{\beta}_{-k}|\otimes\langle\delta{\beta}_{k}|)\nonumber\\
    &+\sum_kq_{k}\left(|\delta c_{k}\rangle\langle\delta{\alpha}|+|\delta{\alpha}\rangle\langle\delta{c_{k}}|\right),
\end{align} 
in which $|\delta c_{k}\rangle$ is the fluctuation state associated with a photon of wave vector $k$, $|\delta\alpha\rangle$ is the fluctuation state of the Kittle magnon, and $|{\delta\beta}_{\pm k}\rangle$ is the fluctuation state of the magnon pair with wave vector $\pm k$.

By the Lippmann-Schwinger formula~\cite{Nonreciprocal,Scattering,Theoretical}, the scattered states of the ``probe" microwave photon read
\begin{align}
    \left|\psi_{ c_{k}}\right\rangle&=\left|\delta c_{k}\right\rangle+\frac{1}{\Omega_{k}-\omega_{d}-\hat{H}_{0}+i 0_{+}} \hat{H}_{\mathrm{int}}\left|\psi_{c_{k}}\right\rangle\nonumber\\
    &=\hat{T}\left|\delta c_{k}\right\rangle,
\end{align} 
which is related to the initial states $\left|\delta c_{k}\right\rangle$ of the microwave photon via the $\hat{T}$-matrix.
The elements of the $\hat{T}$-matrix are derived as~\cite{Nonreciprocal,Many,Scattering,Theoretical} 
\begin{subequations}
\begin{align}
\label{Tk2}
&T_{c_{k^{\prime}},c_{k}}= \langle \delta c_{k^{\prime}}|\hat{T}|\delta c_{k}\rangle\nonumber\\
&= \langle \delta c_{k^{\prime}}\left|\delta c_{k}\right\rangle+\langle \delta c_{k^{\prime}}|\frac{1}{\Omega_{k}-\omega_{d}-\hat{H}_{0}+i 0_{+}}\hat{H}_{\mathrm{int}}\hat{T}|\delta c_{k}\rangle\nonumber
\\&=\delta_{c_{k^{\prime}},c_{k}}+\frac{1}{\Omega_{k}-\Omega_{k^{\prime}}+i 0_{+}} q_{k^\prime} \langle{\delta\alpha}|\hat{T}|\delta c_{k}\rangle,\\
\label{Talpha2}
&T_{\alpha,c_{k}}= \langle{\delta\alpha}|\hat{T}|\delta c_{k}\rangle\nonumber\\
&=\frac{1}{\Omega_{k}-\omega_{0}+i 0_{+}}\Big[g_{k}\langle\hat{\beta}_{k}\rangle\langle\delta{\beta}_{-k}|\hat{T}|\delta c_{k}\rangle\nonumber\\
&+g_{k}\langle\hat{\beta}_{-k}\rangle\langle\delta{\beta}_{k}|\hat{T}|\delta c_{k}\rangle+\sum_{k^{\prime}} q_{k^{\prime }}\langle \delta c_{k^{\prime}}|\hat{T}|\delta c_{k}\rangle\Big],\\
\label{T_beta-k}
&T_{{\beta}_{-k},c_{k}}= \langle{\delta{\beta}_{-k}}|\hat{T}|\delta c_{k}\rangle\nonumber\\ 
&=\frac{1}{\Omega_{k}-\omega_{-k}-{\omega_{d}}/{2}+i 0_{+}}g^{*}_{k}\langle\hat{\beta}^{\dagger}_{k}\rangle\langle{\delta\alpha}|\hat{T}|\delta c_{k}\rangle,\\
\label{T_betak}
&T_{{\beta}_{k},c_{k}}= \langle{\delta{\beta}_{k}}|\hat{T}|\delta c_{k}\rangle\nonumber\\
&=\frac{1}{\Omega_{k}-\omega_{k}-{\omega_{d}}/{2}+i 0_{+}}g^{*}_{k}\langle\hat{\beta}^{\dagger}_{-k}\rangle\langle\delta{\alpha}|\hat{T}|\delta c_{k}\rangle.
\end{align} 
\end{subequations}
By substituting Eq.~\eqref{Tk2}, \eqref{T_beta-k}, and \eqref{T_betak} into \eqref{Talpha2}, we obtain the amplitudes between the probe microwave photon and the fluctuation of the Kittel magnon 
\begin{align}
T_{\alpha,c_{k}}&= \langle\delta{\alpha}|\hat{T}|\delta c_{k}\rangle=\frac{1}{\Omega_{k}-\omega_{0}+i 0_{+}}\nonumber\\
&\times\Big[|g_{k}|^{2}|\langle\hat{\beta}_{k}\rangle|^{2}\frac{1}{\Omega_{k}-\omega_{-k}-{\omega_{d}}/{2}+i 0_{+}}\langle\delta{\alpha}|\hat{T}|\delta c_{k}\rangle\nonumber\\
&+|g_{k}|^{2}|\langle\hat{\beta}_{-k}\rangle|^{2}\frac{1}{\Omega_{k}-\omega_{k}-{\omega_{d}}/{2}+i 0_{+}}\langle\delta{\alpha}|\hat{T}|\delta c_{k}\rangle\nonumber\\
&+\sum_{k^{\prime}}\frac{ q^{2}_{k^{\prime}}}{\Omega_{k}-\Omega_{k^{\prime}}+i 0_{+}} \langle\delta{\alpha}|\hat{T}|\delta c_{k}\rangle+ q_{k}\Big],
\end{align} 
leading to 
\begin{align}
\label{alphak}
T_{\alpha,c_{k}}=\frac{q_{k}}{(\Omega_{k}-\omega_{0}+i 0_{+})-\Sigma(\Omega_{k})},
\end{align} 
in which 
\begin{align}
\Sigma(\Omega_{k})&=\frac{|g_{k}|^{2}|\langle\hat{\beta}_{k}\rangle|^{2}}{\Omega_{k}-\omega_{-k}-{\omega_{d}}/{2}+i 0_{+}}+\frac{|g_{k}|^{2}|\langle\hat{\beta}_{-k}\rangle|^{2}}{\Omega_{k}-\omega_{k}-{\omega_{d}}/{2}+i 0_{+}}\nonumber\\
   &+\sum_{k^{\prime}}\frac{ q^{2}_{k^{\prime}}}{\Omega_{k}-\Omega_{k^{\prime}}+i 0_{+}} 
\end{align}
represents the self-energy of VNA photons~\cite{Many}. Finally, substitution of Eq.~\eqref{alphak} into Eq.~\eqref{Tk2} yields the scattering amplitudes of the ``probe" photons
\begin{align}
&T_{c_{k^{\prime}},c_{k}}=  \langle \delta c_{k^{\prime}}| \hat{T}\left|\delta c_{k}\right\rangle= \delta_{c_{k^{\prime}},c_{k}}+\frac{1}{\Omega_{k}-\Omega_{k^{\prime}}+i 0_{+}}\nonumber\\
&\times\frac{q_{k^{\prime}}q_{k}}{(\Omega_{k}-\omega_{0}+i 0_{+})-\Sigma(\Omega_{k})}.
\end{align}

For the initial state of a probe microwave photon with wave vector $ k> 0$, the scattered state far away from the ``scattering region" magnetic sphere reads 
\begin{align}
    &\langle y | \psi_{c_{k}} \rangle \big|_{y \to +\infty} = \langle y | \hat{T} | \delta c_{k}\rangle\nonumber\\
    &=\sum_{k'} \langle y | \delta c_{k^{\prime}}\rangle \langle \delta c_{k^{\prime}}| \hat{T} |\delta c_{k} \rangle\nonumber\\
    &=\langle y | \delta c_{k}\rangle +\int \frac{dk'}{2\pi} L_{y} \langle y |\delta c_{k^{\prime}} \rangle \frac{1}{\Omega_{k}-\Omega_{k^{\prime}}+i 0_{+}}\nonumber\\
    &\times\frac{q_{k^{\prime}}q_{k}}{[(\Omega_{k}-\omega_{0}+i 0_{+})-\Sigma(\Omega_{k})]}, 
    \label{scattered_state}
\end{align}
in which $L_y$ is the length of the coplanar waveguide along the propagation $\hat{\bf y}$-direction. With the dispersion relation $\Omega_{k}=v_{k}k$ of the photon modes in a waveguide with the photon group velocity $v_k$~\cite{Scattering,Theoretical}, we calculate the integral in \eqref{scattered_state} by applying the residue theorem: 
\begin{align}
    &\int \frac{dk'}{2\pi} L_{y}   \frac{\langle y | \delta c_{k^{\prime}}\rangle}{\Omega_{k}-\Omega_{k^{\prime}}+i 0_{+}}\frac{q_{k^{\prime}}q_{k}}{[(\Omega_{k}-\omega_{0}+i 0_{+})-\Sigma(\Omega_{k})]} \nonumber\\
    &= \frac{q_{k}}{[(\Omega_{k}-\omega_{0}+i 0_{+})-\Sigma(\Omega_{k})]} \left(\frac{-i L_y\langle y |  \delta c_{k} \rangle q_{k} }{v_{k}}\right),
\end{align}
in which 
\begin{align}
\Sigma(\Omega_{k})&=\frac{|g_{k}|^{2}|\langle\hat{\beta}_{k}\rangle|^{2}}{\Omega_{k}-\omega_{-k}-{\omega_{d}}/{2}+i 0_{+}}+\frac{|g_{k}|^{2}|\langle\hat{\beta}_{-k}\rangle|^{2}}{\Omega_{k}-\omega_{k}-{\omega_{d}}/{2}+i 0_{+}}\nonumber\\
&-{i L_yq^{2}_{k} }/{v_{k}},
\end{align}
since $\sum_{k'}{q_{k'}^2}/({\Omega_{k}-\Omega_{k'}+i0^+})=-{i L_yq^{2}_{k} }/{v_{k}}$. Therefore, the scattered state reads
\begin{align}
\label{scattering}
&\langle y | \psi_{c_{k}} \rangle \big|_{y \to +\infty}\nonumber\\
&=\langle y | \delta c_{k} \rangle\left(1-\frac{E_k}{(\Omega_{k}-\omega_{0}+i 0_{+})-F_k-G_k+E_k}\right),
\end{align}
in which 
\begin{align}
&E_k={i L_yq^{2}_{k} }/{v_{k}},\nonumber\\
&F_k=|g_{k}|^{2}|\langle\hat{\beta}_{k}\rangle|^{2}\frac{1}{\Omega_{k}-\omega_{-k}-{\omega_{d}}/{2}+i 0_{+}},\nonumber\\
&G_k=|g_{k}|^{2}|\langle\hat{\beta}_{-k}\rangle|^{2}\frac{1}{\Omega_{k}-\omega_{k}-{\omega_{d}}/{2}+i 0_{+}}.
\end{align}
According to Eq.~\eqref{scattering}, after the microwave signal passes through the coplanar waveguide loaded with the YIG sphere, the incident state $\langle y |  \delta c_{k} \rangle$ is modulated by the microwave transmission.
Including the damping of magnons, the microwave transmission is finally given by 
\begin{align}
S_{21}(\omega)=1-\frac{E_k}{(\Omega_{k}-\omega_{0}+i{\gamma_0}/{2})-F_{\gamma}(k)-G_{\gamma}(k)+E_k},
\label{4S21}
\end{align}
in which, after accounting for the damping, the photon self-energies
\begin{align}
&F_{\gamma}(k)=|g_{k}|^{2}|\langle\hat{\beta}_{k}\rangle|^{2}\frac{1}{\Omega_{k}-\omega_{-k}-{\omega_{d}}/{2}+i{\gamma_{-k}}/{2}},\nonumber\\
&G_{\gamma}(k)=|g_{k}|^{2}|\langle\hat{\beta}_{-k}\rangle|^{2}\frac{1}{\Omega_{k}-\omega_{k}-{\omega_{d}}/{2}+i{\gamma_{k}}/{2}}.
\end{align}

\section{\label{calculation}Comparison of observations and model calculations}

\begin{figure}[htbp]
    \centering
    \includegraphics[width=0.87\linewidth]{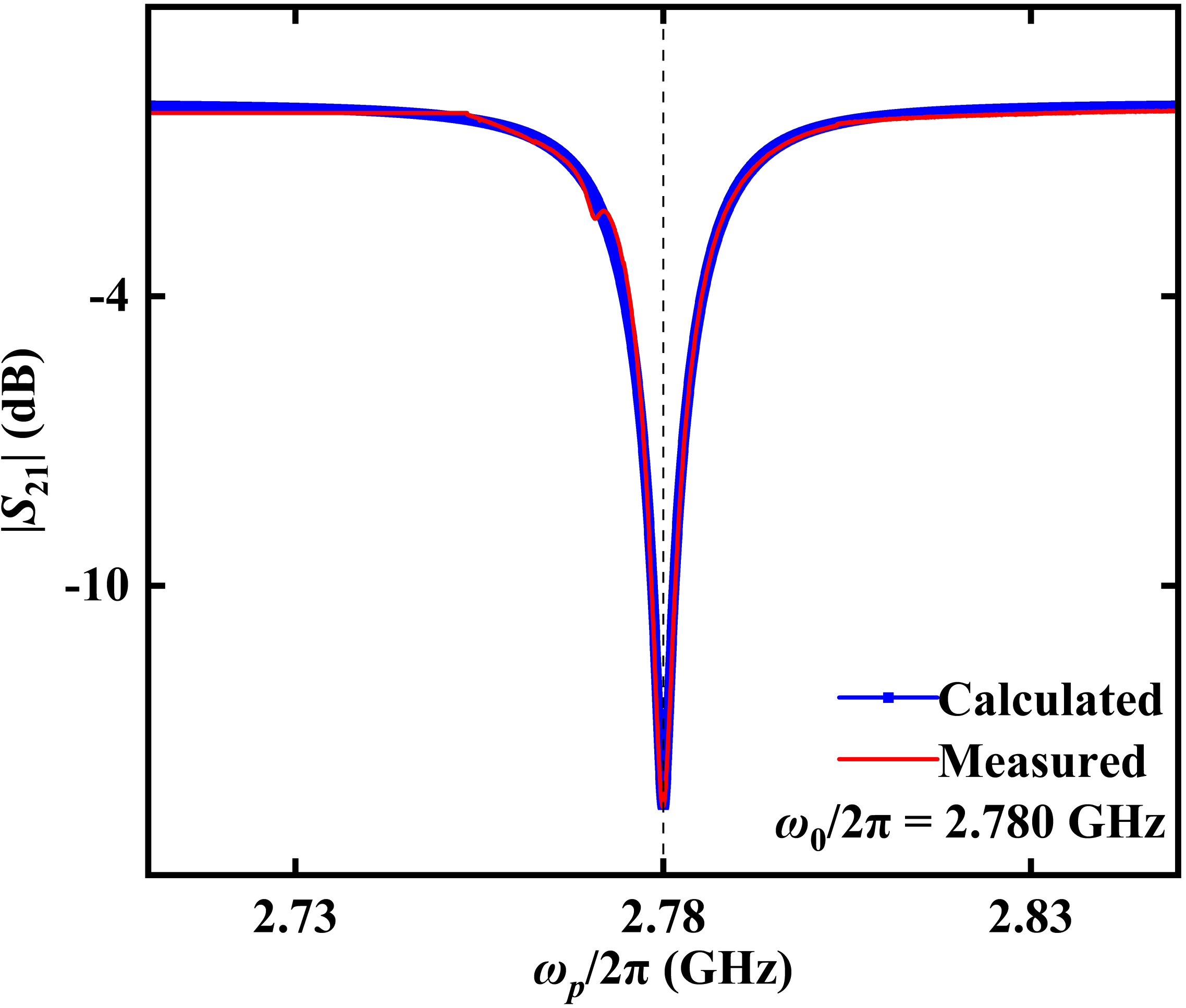}
    \caption{Comparison of measured FMR absorption spectra and calculated microwave transmission. The red curve is the measured FMR absorption spectrum of the Kittel magnon mode with the external magnetic field $\mu_0H_{\rm ext}=92.265$~mT. The blue curve represents a fit of the microwave transmission by Eq.~\eqref{4S21} to the experimental data, yielding $\omega_0/2\pi=2.780$~GHz and $\gamma_0/2\pi=1.5$~MHz.}
    \label{Fig.7}
\end{figure}

To obtain the intrinsic parameters of the photon modes in the coplanar waveguide and the FMR in the magnetic sphere, we first consider the coupling between the Kittel magnon mode and the waveguide photon mode without exerting the drive microwaves, in which case only the FMR absorption spectrum exists.
In this linear-response regime, we disregard the photon self-energies $F_{\gamma}(k)$ and $G_{\gamma}(k)$ due to the three-magnon interaction in Eq.~\eqref{4S21} when calculating the FMR absorption spectrum.
We use the parameters $v_k=3 \times 10^7$~m/s, $q_k=40$~MHz, and $\gamma_0/2\pi=1.5$~MHz in the calculation Eq.~\eqref{4S21}. Notably, this coupling strength $q_k$ corresponds to a radiative damping of $|E_k| \approx 2\pi\times 8.48$~MHz, which agrees with the typical radiative damping of FMR of a YIG sphere in microwave waveguides~\cite{Chiral,radiative}.
We find that the calculated transmission spectrum well reproduces the experimentally measured FMR spectrum, as compared in Fig.~\ref{Fig.7}.

\begin{figure}[htbp]
    \centering
    \includegraphics[width=0.98\linewidth]{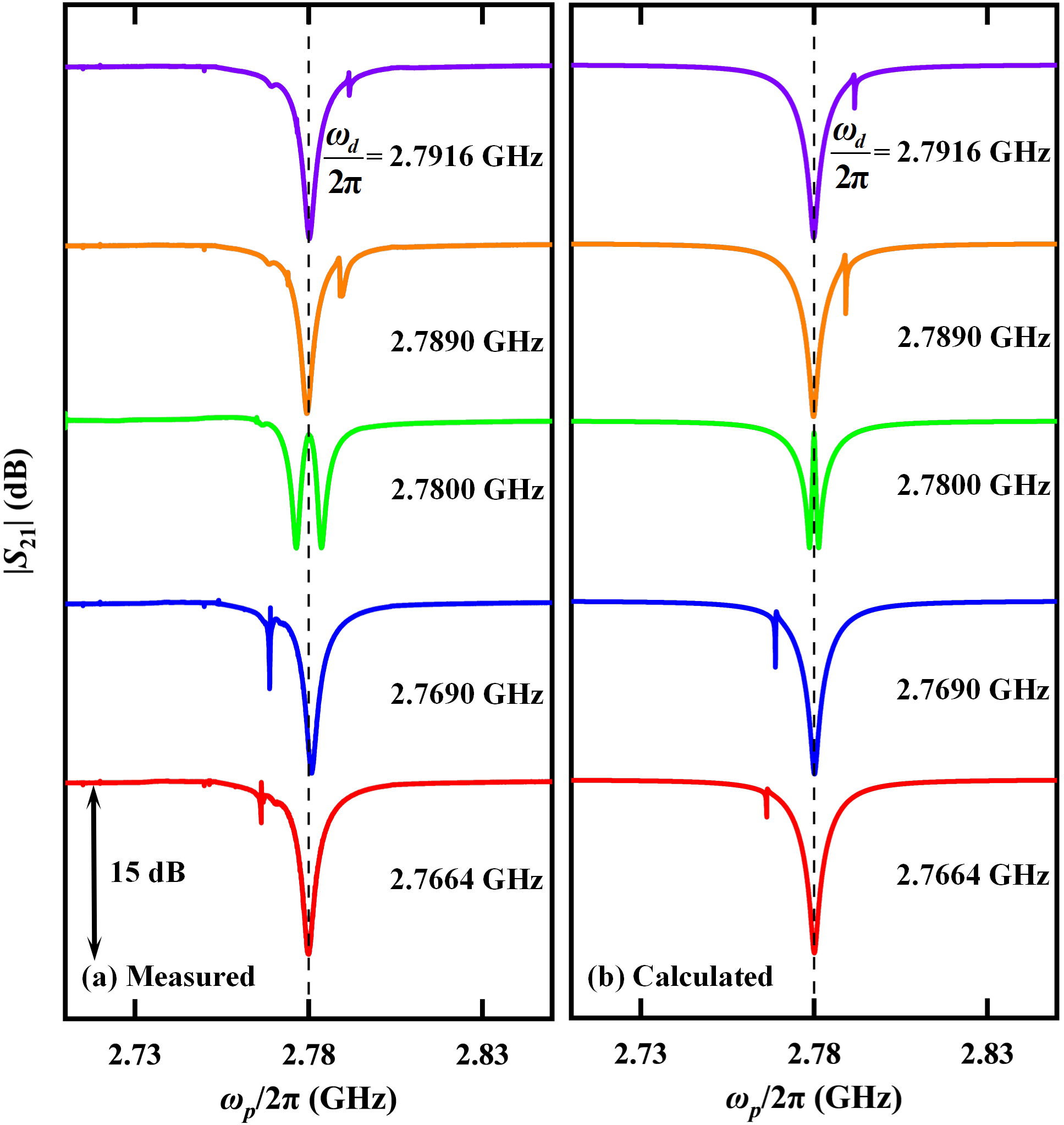}
    \caption{Comparison of experimental measurements and theoretical calculations for the microwave transmission spectra with different pump frequencies. (a) Experimental measurement with typical driven frequencies around the FMR frequency $\omega_0/2\pi=2.780$~GHz. (b) Theoretical calculation of the microwave transmission with the same drive frequencies as in (a). Theoretical calculations can reproduce the observed characteristic Fano resonance and mode splitting.}
    \label{Fig.8}
\end{figure}

To explain the observed Fano resonance occurring around the pump frequencies, as well as the mode splitting, in the presence of strong microwave drive, we include the photon self-energies $F_{\gamma}(k)$ and $G_{\gamma}(k)$ in Eq.~\eqref{4S21} while retaining the parameters obtained in Fig.~\ref{Fig.7}. 
In the calculation, we use the damping broadening $\gamma_k/2\pi=\gamma_{-k}/2\pi=0.11$~MHz for the magnon pair, which is much smaller than the FMR linewidth $\gamma_0$, $\gamma_{\rm ext}/2\pi=0.5$~MHz, $\omega_k=\omega_{-k}={\omega_d}/2$, and $g_{k}=3$~Hz.
The steady-state value of $|\langle\hat{\beta}_{k}\rangle| =|\langle\hat{\beta}_{-k}\rangle|$ is solved according to Eq.~\eqref{equation_beta} for different driven frequencies $\omega_{d}$. 
Here we present their corresponding values of $|\langle\hat{\beta}_{k}\rangle|=\{1.429, 1.879, 1.924, 1.879, 1.429\}\times10^6$ when $P_{d}=-5$~dBm and $\omega_{d}/2\pi= \{2.766, 2.775, 2.780, 2.785, 2.794\}$~GHz. In this case, the effective coupling between $\delta\alpha$ and $\delta\hat{\beta}^{\dagger}_{\pm k}$ is about 6~MHz. The value of $|\langle\hat{\beta}_{k}\rangle|$ reaches its maximum $\beta$ [Eq.~\eqref{beta}] when $\omega_{d}=\omega_{0}$, but the variations in numerical values are not significant when $\omega_d$ is around $\omega_0$. 
We compare five characteristic microwave transmission spectra at different pump frequencies in Fig.~\ref{Fig.8}. 
Figure~\ref{Fig.8}(a) shows the measurement, while Fig.~\ref{Fig.8}(b) is obtained by substitution of different values of $\omega_d$ into Eq.~\eqref{4S21}.

As shown in Fig.~\ref{Fig.8}, the calculated transmission spectrum well reproduces the features of the Fano resonance when $\omega_d$ is a bit away from the FMR $\omega_0$ as well as the mode splitting when $\omega_d=\omega_0$. 
When the pump frequency $\omega_d$ approaches the FMR frequency $\omega_0$, asymmetric Fano-resonance shapes appear in the transmission spectra, manifesting as a combination of sharp peaks and dips. To understand the appearance of the Fano resonance shape, we analyze the denominator $D=(\Omega_{k}-\omega_{0}+i{\gamma_0}/{2})-F_{\gamma}(k)-G_{\gamma}(k)+E_k$ in the microwave transmission Eq.~\eqref{4S21}. 
To this end, we decompose the real and imaginary parts of $D$ according to  
\begin{align}
\label{reim}
&{\rm Re}[D]=(\Omega_{k}-\omega_{0})-2{\rm Re}[F_{\gamma}(k)],\nonumber\\
&{\rm Im}[D]={\gamma_{0}}/{2}-2{\rm Im}[F_{\gamma}(k)]+|E_{k}|,
\end{align}
in which 
\begin{align}
\label{FG}
&{\rm Re}[F_{\gamma}(k)]={\rm Re}[G_{\gamma}(k)]=|g_{k}|^{2}|\langle\hat{\beta}_{k}\rangle|^{2}\frac{\Omega_{k}-{\omega_{d}}}{(\Omega_{k}-{\omega_{d}})^2+(\frac{\gamma_{k}}{2})^2},\nonumber\\
&{\rm Im}[F_{\gamma}(k)]={\rm Im}[G_{\gamma}(k)]=-|g_{k}|^{2}|\langle\hat{\beta}_{k}\rangle|^{2}\frac{{\gamma_{k}}/{2}}{(\Omega_{k}-{\omega_{d}})^2+(\frac{\gamma_{k}}{2})^2}.
\end{align}
Combining Eqs.~\eqref{reim} and \eqref{FG}, we obtain 
\begin{align}
&{\rm Re}[S_{21}]=1-|E_{k}|{\rm Im}[D]/|D|^2,\nonumber\\
&{\rm Im}[S_{21}]=-|E_{k}|{\rm Re}[D]/|D|^2.
\label{absS21}
\end{align}
According to Eqs.~\eqref{reim} and \eqref{FG}, when the frequency of the ``probe" photon coincides with the driven frequency, i.e., $\Omega_{k}=\omega_{k}+{\omega_{d}}/{2}=\omega_{d}$, ${\rm Re}[F_{\gamma}(k)]=0$ and ${\rm Re}[D]=\omega_d-\omega_0<\gamma_0$ is small, rendering ${\rm Im}[D]$ to dominate in the microwave transmission spectrum.
Since $\gamma_{k}$ is sufficiently small, ${\rm Im}[F_{\gamma}(k)]$ is a large component when $|\Omega_{k}-\omega_d|\approx \gamma_k/2$ in ${\rm Im}[D]$. Particularly, when $\Omega_{k}=\omega_{d}$, $|{\rm Im}[D]|$ reaches its maximum, and the transmission spectrum $|S_{21}| \rm(dB)=20\times log_{10}|S_{21}|$, therefore, exhibits a peak \textit{exactly} at $\Omega_{k}=\omega_d$ (refer to, for example, Fig.~\ref{Fig.9} below).

According to the measurement, the distribution of the asymmetric spectra, i.e., dip followed by peak or peak followed by dip, depends on whether $\omega_{d}< \omega_{0}$ or $\omega_{d}>\omega_{0}$. This feature is well captured by our scattering theory of photons, which accounts for the three-magnon interaction. 
We first analyze the situation when $\omega_{d}< \omega_{0}$, with which the Fano resonance exhibits a dip followed by a peak. 
Continuing the analysis of Eq.~\eqref{reim}, we observe that when $\Omega_{k}\neq\omega_{d}$, the imaginary component ${\rm Im}[D]$ decreases with the increase of $|\Omega_{k}-\omega_{d}|$, which is a cause of the reduction in ${|D|}$. At the same time, when $\Omega_{k} <\omega_{d}<\omega_{0}$, $\Omega_{k}-\omega_{0}$ and $-2{\rm Re}[F_{\gamma}(k)]$ in the real component ${\rm Re}[D]$ are opposite in sign and cancel each other, resulting in a minimum appearing in ${\rm Re}[D]$ when $\Omega_k\rightarrow \omega_d$. 
Accordingly, from Eq.~\eqref{absS21}, $|S_{21}|$~(dB) is proportional to ${|D|}$, so when $\Omega_k\rightarrow \omega_0$ the transmission spectra first exhibits a dip at $\Omega_k<\omega_d$ and a peak at $\Omega_k=\omega_d$.  Similarly, when analyzing the case with $ \omega_d>\omega_{0}$, we conclude that a dip appears at $\Omega_k>\omega_d$ and a peak at $\Omega_k=\omega_d$.

At the ferromagnetic resonance with $\omega_d=\omega_0$, two dips of equal intensity appear on both sides of $\omega_0$, with the positions corresponding to the real component of two complex roots $\omega_{1,2}$ in 
\[
D(\Omega_k\rightarrow \omega_{1,2})=0,
\]
while the imaginary part is for their frequency broadening. Solving the equation $D(\omega)=0$ yields two roots $\omega=\omega_{1,2}$; their real part corresponds to a frequency splitting $\Delta_{\rm{splitting}} = {\rm Re}[\omega_1] - {\rm Re}[\omega_2] \approx 2\sqrt{2}|g_k||\langle\hat{\beta}_k\rangle|$. Substitution of $\beta \approx \sqrt{\Omega_d/|g_k|}$ from Eq.~(8) reveals that the splitting strength scales as $\Delta_{\mathrm{splitting}} \propto \sqrt{|g_k|\Omega_d}\propto P_d^{1/4}$, which is explicitly governed by the coupling strength $g_k$ and the driven power $P_d$. Substitution ${\rm Re}[\omega_{1,2}]$ into Eq.~\eqref{reim} reveals that when $\omega_d=\omega_0$, ${\rm Re}[D]$ and ${\rm Im}[D]$ are respectively odd and even functions of ${\rm Re}[\omega_{1,2}]-\omega_0$. Consequently, the transmission spectrum $|S_{21}|$~(dB) exhibits a symmetrical distribution about $\omega_0$.

It is noted that focusing on the three-magnon interaction process is an approximation since many other modes may be involved in the nonlinear four-magnon interaction~\citep{Direct,Compound}. These interactions may influence the mode damping and the dynamics of the fluctuation. Despite these, the quantitative features appear to be well reproduced by the simplified model. 

We measure the microwave transmission spectra at different bias magnetic fields, corresponding to the FMR frequencies $\omega_0/2\pi=\{1.980,2.280,2.580,2.780\}$~GHz, and find the associated damping rates $\gamma_k/2\pi=\{0.07,0.15,0.17,0.11\}$~MHz of magnon pairs all correspond to a long lifetime.

Finally, we analyze the effect of the pump power on the Fano resonance. We compare the experimental measurement and theoretical calculation of the Fano resonance at different powers for two typical drive frequencies $\omega_{d}/2\pi=2.770$~GHz and $2.788$~GHz in Fig.~\ref{Fig.9}.
Comparing Fig.~\ref{Fig.9}(a,c) with (b,d), we observe that the theoretical calculations can well capture the characteristics of the Fano resonance at different pump powers. As the power increases, both the peaks and dips in the Fano resonance become more pronounced, and the resonance region expands. This is because the coupling between the driven microwave photon and the Kittel magnon $\Omega_d\propto \sqrt{P_{d}}$, such that the photon self-energy $\{F_{\gamma}(k),G_{\gamma}(k)\} \propto \beta^2\propto \sqrt{P_{d}}$ [refer to Eq.~\eqref{beta}] increase as the pump power increases, rendering the Fano resonance phenomenon more pronounced.

\begin{figure}[htp!]
    \centering
    \includegraphics[width=1\linewidth]{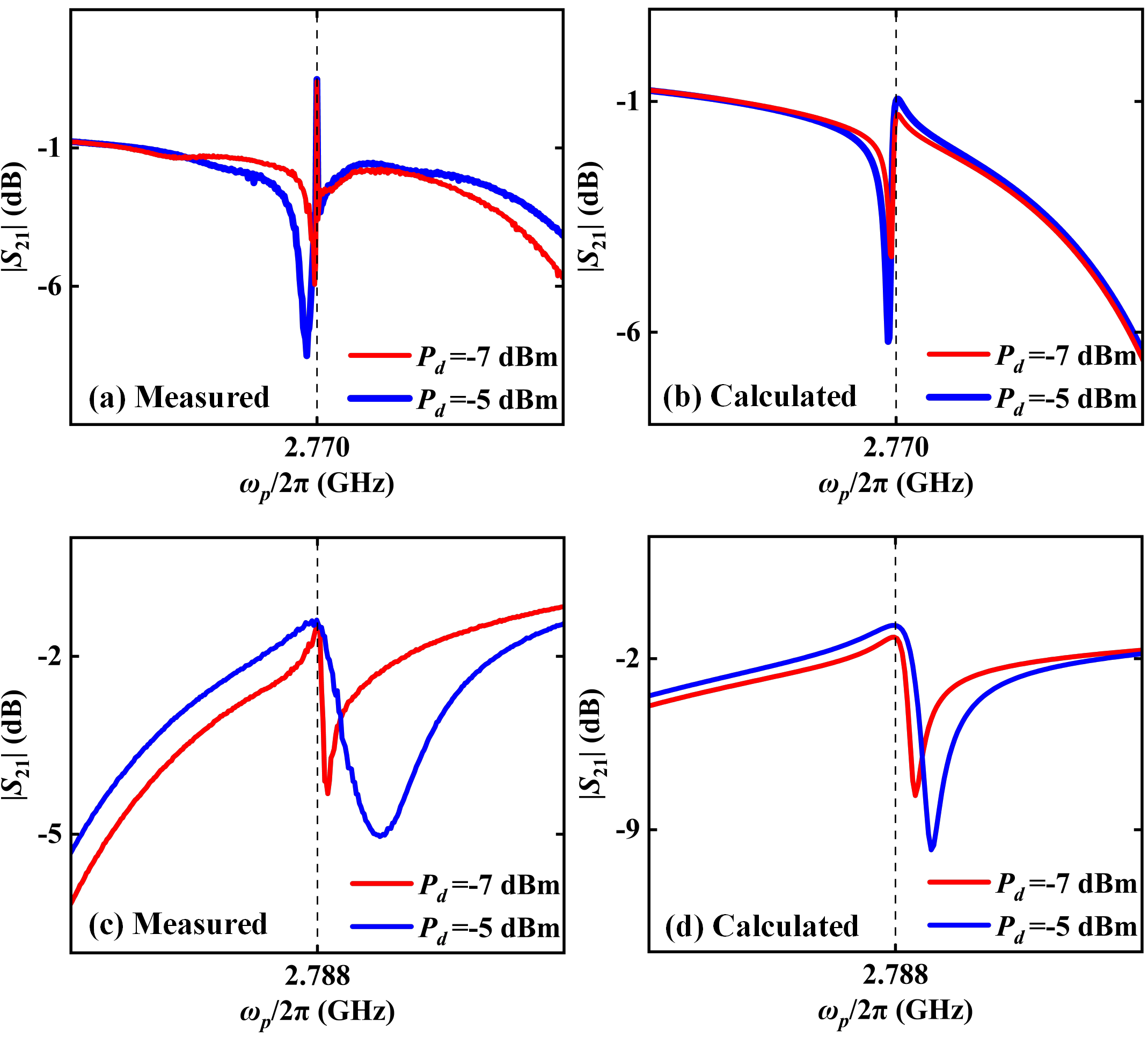}
    \caption{Comparison of typical experimental measurements and theoretical calculations of the microwave transmission spectra at different pump powers. (a) and (c) Microwave transmission spectra measured experimentally at different pump powers. (b) and (d) Calculated microwave transmission spectra at different pump powers.}
    \label{Fig.9}
\end{figure}

\section{\label{summary}Discussion and conclusion}

Fano resonance has been explored in various physical systems, such as metasurfaces\cite{Metasurfaces}, subwavelength gratings\cite{Momentum,Abrupt}, photonic crystals~\citep{Plasmon,Mie,photonic,Enhanced}, metamaterials~\citep{Sharp,resonant,metamolecule}, and plasmonic systems~\citep{Light,Influence,nanostructures,Quantum}, which holds potential application across a wide range of research fields, from telecommunications to hypersensitive biosensing, medical instruments, and data storage. Specifically, in hybrid magnonics, Fano resonances provide a powerful tool for revealing coupling mechanisms in magnon-photon and magnon-phonon nonlinear interactions since the damping of photons and phonons is much smaller than the magnons~\citep{Cavity,Coherent,multiwindow,Measuring,Generated}, as well as detecting magnon-qubit coupling~\cite{magnonqubit}. However, it is unexpected that the Fano resonance can appear in the nonlinear magnon interactions since the lifetimes of different magnons are usually comparable. We demonstrate in this work the merit of Fano resonance for detecting nonlinear interactions \textit{among magnons}.

In conclusion, we observe a typical Fano resonance due to the nonlinear magnetization dynamics in the microwave transmission spectra as the pump frequency of the drive microwaves approaches, but does not coincide with, the ferromagnetic resonance. 
As the pumping frequency is very close to the ferromagnetic resonance, a mode splitting occurs. By constructing a scattering theory of microwave photons taking into account the three-magnon interaction, we interpret that these phenomena originate from the coupling between the Kittel magnon and a pair of magnons with frequency $\omega_d/2$, holding opposite wave vectors. 
Based on this model, we reveal that the microwave transmission measures the dynamics of the fluctuations $\delta \hat{\alpha}$ of the Kittel magnon and $\delta\hat{\beta}_{\pm k}$ of the magnon pair of frequency $\omega_d/2$, which are coupled due to the driven steady-state amplitudes of the Kittel magnon and magnon pair.
Theoretical calculations reproduce the observed features well, indicating that the occurrence of the Fano resonance is attributable to the damping of $\delta\hat{\beta}_{\pm k}$ being significantly smaller than that of $\delta \hat{\alpha}$. Our research may inspire microwave engineering and the potential application of low-dissipation magnon pairs in future quantum and classical information processing.

\begin{acknowledgments}
This work is financially supported by the National Key Research and Development Program of China under Grant No.~2023YFA1406600 and the National Natural Science Foundation of China under Grant No.~12374109. We thank Zhiyuan Sun and Jinwei Rao for the inspiring discussion. 
\end{acknowledgments}

\section*{Data Availability Statement}
The data that support the findings of this study are available from the corresponding author upon reasonable request.

\nocite{*}

\bibliography{aipsamp}

\end{document}